\documentclass[preprintnumbers,numbers,sort&compress,
               nofootinbib,3p,twocolumn,
                            showpacs,
               colorlinks,
               linkcolor=blue,
               citecolor=blue]{elsarticle}

\def\ra{\rangle}
\def\la{\langle}

\usepackage{graphicx,amsmath,amssymb,bm}
\usepackage{psfrag}
 \usepackage{feynmp}
\usepackage{hyperref}
\usepackage{enumitem}
  \newcommand{\exclude}[1]{}
\newcommand{\be}{\begin{eqnarray}}
\newcommand{\ee}{\end{eqnarray}}

\begin{document}
\title{The 21cm Absorption Line and the Axion Quark Nugget Dark Matter Model } 
\author{K. Lawson and A.R. Zhitnitsky}
\address{ Department of Physics \& Astronomy, University of British Columbia, 
Vancouver, B.C. V6T 1Z1, Canada} 

\begin{abstract} 
We argue that dark matter in the form of macroscopically large nuggets of  
standard model quarks and antiquarks can help to alleviate the tension between 
standard model cosmology and the recent EDGES observation of a stronger than 
anticipated 21 cm absorption feature.  The effect occurs as a result of the  thermal 
emission from quark nugget dark matter at early times and at energies well below 
the peak of the CMB. Similar radiation at early times may also contribute 
a fraction of the GHz range excess observed by ARCADE2. 
\end{abstract}

\maketitle

\section{Introduction}
Redshifted 21 cm emission serves as a tracer of baryonic physics in the early 
universe prior to reionization. Observations by the low-band antenna of the 
Experiment to Detect the Global EoR Signal (EDGES) of a stronger 
than anticipated 21 cm absorption feature pose a challenge to our standard 
picture of the early universe \cite{Bowman:2018yin}. 
This signal suggests a larger than expected difference between the 
radiation temperature and the spin temperature of the absorbing hydrogen gas. In 
particular it has been suggested that a stronger than expected radio background present  
at $z>17$ may be capable of producing the EDGES result \cite{Feng:2018rje}. 
The possibility of a low energy excess was also indicated by the earlier observations of 
ARCADE2 in the GHz range \cite{Fixsen:2009xn} but at a level significantly above that 
required to produce the EDGES feature. 

The EDGES result has sparked a number of proposals designed to alleviate 
its tension with the standard cosmological framework. In particular,
it has been suggested that  the 
dark matter (DM) may act as a heat sink cooling the neutral gas relative to the CMB 
\cite{Munoz:2018pzp, Barkana:2018qrx, Fialkov:2018xre, Slatyer:2018aqg, Li:2018kzs}. 
However this  idea seems to be subject to several very strong 
constraints \cite{Berlin:2018sjs, Munoz:2018jwq, Mahdawi:2018euy, Kovetz:2018zan, Neufeld:2018slx} 
and may be inconsistent with results from the CMB and Lyman-alpha flux 
\cite{Xu:2018efh}.

Another proposal to ease the tension with the conventional model is to  introduce an additional 
source of soft radiation at large redshifts ($z\gtrsim 17$) as suggested in \cite{Fraser:2018acy} 
using DM particles \cite{Pospelov:2018kdh, Clark:2018ghm} 
or intermediate mass black holes as suggested in 
\cite{Ewall-Wice:2018bzf}.  However, models in which dark matter or 
black holes are capable of modifying  the soft photon spectrum generally also modify the hard 
photon spectrum, which is strongly constrained by independent observations. Therefore, 
any model introduced to explain the 
EDGES result must produce emission which is strongly suppressed 
at energies above the radio band. Additionally, synchrotron production of the required 
soft photon background is disfavoured by the relatively short cooling time of 
synchrotron emitting astrophysical electrons \cite{Sharma:2018agu}.
Alternatively it has been suggested that a different choice of foreground parameters than 
those used in \cite{Bowman:2018yin} may give very different properties of the 
absorption profile \cite{Hills:2018vyr}.
 
Here we present a proposal which may remove the tensions 
related to the EDGES results as highlighted above. 
The arguments are based on the Axion Quark Nugget (AQN) 
dark matter model. This model was originally invented as an  
explanation of  the observed similarity in the visible and dark matter densities 
\be
\label{eq:dark-vis}
\Omega_{\rm dark} \sim   \Omega_{\rm visible} 
\ee
as it naturally produces theses components at the same scale  
independent of the specific values of any of the parameters of the model. 
We overview the basic ideas of this model in section \ref{sec:AQN} but 
first make two important comments related to this model in the context of the present work. 

First, the AQN model  was invented long ago \cite{Zhitnitsky:2002qa}, 
though a specific formation mechanism  for the nuggets has been 
developed in the much more recent papers \cite{Liang:2016tqc,Ge:2017ttc,Ge:2017idw}. 
It was proposed independently of the recent EDGES observations \cite{Bowman:2018yin} 
and, in contrast  
with much recent activity in the field, the AQN model was not invented specifically to 
explain the EDGES result. Rather, this model was invented  
as a natural explanation of the observed dark matter to baryon ratio 
of expression (\ref{eq:dark-vis}).
The similarity between the dark matter $ \Omega_{\rm dark}$ and the visible matter 
$\Omega_{\rm visible}$  densities strongly suggests that both types of matter  
formed  during the same cosmological epoch. For the baryon density this is 
obviously the QCD phase transition as the baryon mass $m_p$ is is   
proportional to $ \Lambda_{\rm QCD}$ while the electroweak contribution 
proportional to the quark masses $m_q$ represents only a
minor contribution to the baryon mass.  

The second comment we would like to make in this Introduction is as follows. 
Observations made by ARCADE2 in the GHz range \cite{Fixsen:2009xn} 
hinted at a possible excess of diffuse radiation. The authors of \cite{Feng:2018rje}
argue that just $\sim 1\%$ of this excess is needed to explain 
the EDGES result. We have previously argued in \cite{Lawson:2012zu} that the AQN model will 
necessarily contribute to the radio background across a broad range from the MHz up 
to the GHz scale, though a precise computations within AQN model are hard to carry out due to very high
sensitivity to some unknown parameters. However, the main  conclusion  of  \cite{Lawson:2012zu}
holds in a sense that  the AQN model offers a potential mechanism to generate the soft photon 
background assumed to be present in \cite{Feng:2018rje} and capable of producing 
the EDGES result. 

At this time the EDGES result seems to indicate that the dominant contribution to  ARCADE2 
cannot be associated with any excess prior to $z\sim 20$ as it would overproduce the observed 
absorption feature. Instead, only a small portion 
about 1\% of the ARCADE2 excess  is consistent with EDGES result, which we assume to be the case.

Our presentation is organized as follows. In Section \ref{sec:AQN} we provide an 
overview of the AQN dark matter model with emphasis on cosmological and astrophysical 
implications. In section \ref{thermal} we discuss some specific features of the spectrum  
radiated by the nuggets at large $z$. In Section \ref{sky-temperature} we compute the resulting 
sky temperature and argue that recent EDGES observations can be understood 
within the AQN framework. 

\section{Axion Quark Nugget dark matter model}\label{sec:AQN}

The idea that the dark matter may take the form of composite objects composed of 
standard model quarks in a novel phase goes back to quark nuggets  \cite{Witten:1984rs}, 
strangelets \cite{Farhi:1984qu} and nuclearities \cite{DeRujula:1984axn}, for a large number 
of references to these original results see the review \cite{Madsen:1998uh}. 
In the models of \cite{Witten:1984rs,Farhi:1984qu,DeRujula:1984axn,Madsen:1998uh}  
the presence of strange quarks stabilizes quark matter at sufficiently 
high densities allowing strangelets formed in the early universe to remain stable 
over cosmological timescales.    There were however a number of problems with the original 
model which shall not be discussed here\footnote{\label{first-order}In particular, a first order 
phase transition is required for the strangelets to be formed during the QCD phase transition.  
However recent lattice results \cite{Aoki:2006we} unambiguously show that the QCD transition 
is a crossover rather than a first order phase transition. Furthermore, the strangelets 
will likely evaporate on the Hubble time-scale even if they had been formed \cite{Alcock:1985}.}.

The AQN model \cite{Zhitnitsky:2002qa} is so named in order to emphasize the essential 
role of the axion field in its construction and to avoid confusion with the earlier models    
\cite{Witten:1984rs,Farhi:1984qu,DeRujula:1984axn,Madsen:1998uh} mentioned above.
It  is drastically different from the previous proposals in two key aspects:\\
1. There is an  additional stabilization factor in the AQN  model provided    
by the axion domain walls which are copiously produced during the QCD transition, 
such that a first order QCD phase transition as mentioned in footnote \ref{first-order} is 
not required.\\
2. The AQN  could be made of matter as well as {\it antimatter} in this framework 
as a result of separation of charges.  Precisely this feature plays a key role in the 
context of the present work. 
  
The basic idea of  the AQN  proposal can be explained   as follows: 
It is commonly  assumed that the Universe 
began in a symmetric state with zero global baryonic charge 
and later (through some baryon number violating process, the so-called baryogenesis) 
evolved into a state with a net positive baryon number. As an 
alternative to this scenario we advocate a model in which 
``baryogenesis'' is actually a charge separation process 
in which the global baryon number of the Universe remains 
zero. The baryon to photon ratio $\eta=n_B/n_{\gamma}\sim 10^{-10}$
is determined by the formation temperature $T_{\rm form}\simeq 40 $ MeV when 
the nuggets complete their formation. The AQN  framework is very different from  
the conventional baryogenesis mechanism in which a single extra baryon must be 
produced for every $10^{10}$ particles in the primordial plasma. 

In this model the unobserved baryons and antibaryons locked in the nuggets come to comprise 
the dark matter in the form of quarks and antiquarks in a  dense 
colour superconducting (CS) phase.  The formation of the  nuggets made of 
matter and antimatter occurs through the dynamics of shrinking axion domain walls. 
For technical details of the formation process which will only be briefly 
discussed here see the recent original papers \cite{Liang:2016tqc,Ge:2017ttc,Ge:2017idw}. 

The nuggets are originally compressed by the collapse of an axion domain wall 
at the time of the QCD   transition. With the domain wall collapsing until 
it is countered by the fermi pressure of the quarks inside. Consequently they have 
a typical size $R\sim (10^{-5}-10^{-4})$ cm determined by the axion mass 
$m_a$ with $R\sim m_a^{-1}$.
It is important to emphasize that there are strong constraints on the allowed window 
for the axion mass,  which can be represented as follows 
$10^{-6} {\rm eV}\leq m_a \leq 10^{-2} {\rm eV}$, see 
recent reviews \cite{vanBibber:2006rb, Asztalos:2006kz,Sikivie:2008,Raffelt:2006cw,
Sikivie:2009fv,Rosenberg:2015kxa,Marsh:2015xka,Graham:2015ouw,Ringwald:2016yge} 
on the subject. The Compton wavelength of the axion represents that largest scale in the 
AQN formation process and consequently sets a maximum baryon number for the 
nuggets $B^{1/3}\lesssim \frac{m_{\pi}}{m_a}$ \cite{Ge:2017idw} strongly limiting the 
parameter space in which the AQN model may be stable and comprise the dark matter. 
It is a highly nontrivial self-consistency check of this model that the axions search constraints 
on $m_a$ and independent  constraints on the mass of the nuggets ${\cal{M}}\simeq B m_p$ 
are overlapping as we review below. 
 
At temperatures below the QCD phase transition the quarks (in the nuggets) or 
antiquarks (in the antinuggets)  inside a closed axion wall cool and settle into a high 
density colour superconducting phase \cite{Liang:2016tqc}. It is the 
degeneracy pressure of this state which supports the domain wall against further collapse. 
Under the conditions considered here this pressure is significantly larger than the thermal 
pressure inside the domain wall. 
At asymptotically large densities the ground state would contain an equal number of u, d and 
s quarks (the so called CFL phase) while at lower densities s quarks are relatively depleted 
due to their larger mass. The resulting combination of u and d quarks is electrically charged, 
positive in the case of quark matter and negative in the case of antiquarks. There is thus a 
strong electrical potential surrounding the nuggets which rapidly results in the formation of 
a surrounding layer of leptons (captured from the surrounding primordial plasma) electrons 
in the case of a quark nugget and positrons in the case of an antiquark nugget. This lepton 
layer is known as the electrosphere and largely determines the observable properties of 
the nuggets. The emergence of the electrosphere  characterized by the chemical potential 
$\mu$ is not a specific feature of the AQNs, rather it is a generic property of all quark nugget 
models \cite{Witten:1984rs,Farhi:1984qu,DeRujula:1984axn,Madsen:1998uh}. 
While the details of the interaction between the electrosphere and the surrounding plasma 
are highly complicated the majority of the leptons are in tightly bound $\mu \sim 10$MeV 
states so that their interactions are suppressed when $T<1$MeV.
The electrosphere is capable of emitting photons with sufficient efficiency that nuggets may 
be cold relative to the background plasma for much of their existence. 
For further details on the low temperature structure of the electrosphere when the core 
of the nuggets is in a CS phase see \cite{Forbes:2009wg}.

The allowed mass of the AQN is constrained by a variety of direct and indirect 
observational constraints. As the AQN are strongly interacting searches constrain   
their number density rather than interaction strength, consequently they impose 
constraints on,
\begin{equation}
n_{N} = \frac{\rho_{DM}}{\la{\cal M}\ra} \approx   \frac{\rho_{DM}}{m_p \la B \ra}
\end{equation}
where $\la{\cal M}\ra$ is the average mass of the nuggets and 
$\la B \ra$ is their average baryon number, not to be confused with the baryon number 
$B$  of individual nuggets, see (\ref{B-range-flares}) below. The low mass regime is strongly 
constrained by a variety of direct searches \cite{Jacobs:2014yca, Lawson:2013bya} requiring
\begin{equation} 
\label{B-range}
\la{\cal M}\ra \gtrsim10^{-2} {\rm{kg}} , ~~~~
|\la B \ra | \gtrsim 10^{25},
\end{equation}
while some individual nuggets could have much lower baryon charge $B\ll 10^{25}$ as discussed below.

Lensing constraints apply at higher masses but only above the region favoured 
by the allowed axion mass range\cite{Jacobs:2014yca}. Lunar seismology also limits the fraction 
of the nugget population that falls in the mass range 10 kg - 1000 kg 
($10^{28} \lesssim |B| \lesssim 10^{30}$) to account for less than $10\%$ of the 
local dark matter flux \cite{Herrin:2005kb}. This result is however sensitive to  the 
efficiency with which the nuggets couple to long wavelength seismic waves. Limits 
on the contribution of antiquark nugget annihilations to earth heating also limit the 
the mean baryon charge of the nuggets to $\la B \ra > 3\times 10^{24}$ \cite{Gorham:2012hy}, 
consistent with the constraint (\ref{B-range}). 

This model is perfectly consistent with all known astrophysical, cosmological, satellite 
and ground based constraints within the parametrical range for 
the mass $\la {\cal M}\ra $ and the baryon charge $\la B\ra$ given by (\ref{B-range}). It is also consistent 
with known constraints from the axion search experiments. Furthermore, there are a number of 
frequency bands where some excess of emission was observed, but not fully explained 
by conventional astrophysical sources. Our comment here is that this model may explain 
some portion, or even the entire excess of the observed radiation in these frequency bands, 
see the short review \cite{Lawson:2013bya} and additional references at the end of this section. 

Another key element of this model is the coherent axion field $\theta$ which is assumed to be 
non-zero during the early Universe QCD transition.
As a result of these $\cal CP$ violating and non-equilibrium processes 
the number of nuggets and antinuggets 
formed will be different. This difference is always an order one effect   
\cite{Liang:2016tqc,Ge:2017ttc,Ge:2017idw} irrespective of the parameters of the theory, 
the axion mass $m_a$ or the initial misalignment angle $\theta_0$. This 
disparity between nugget and antinugget formation produces a similar disparity  
between visible quarks and antiquarks.  This  is precisely  the reason the resulting 
visible and dark matter densities must be the same order of magnitude: 
they are both proportional to the same fundamental $\Lambda_{\rm QCD} $ scale   
and they both are originated at the same QCD epoch resulting in the relation in 
expression (\ref{eq:dark-vis}). If these processes 
are not fundamentally related then the two components $\Omega_{\rm dark}$ and 
$\Omega_{\rm visible}$  could easily exist at vastly different scales. 
 
If we assume that the quark nuggets explain the entirety of both the 
baryon asymmetry and the dark matter then the matter component of the universe consists 
of antiquark nuggets with mass density $\rho_{\bar{N}}$, quark nuggets with mass density 
$\rho_{N}$ and free baryons with mass density $\rho_B$ in the approximate ratio 
\begin{equation}
\label{ratio}
\rho_{\bar{N}}:\rho_N:\rho_B\simeq 3:2:1,
\end{equation}
which corresponds to a zero total baryon charge of the Universe, 
while  the ratio of the visible to DM densities $\rho_{DM}=(\rho_{N}+\rho_{\bar{N}})$ 
takes the observed value of 
$\rho_B/(\rho_{N}+\rho_{\bar{N}})\sim 1/5$. For a more  quantitative treatment of this 
problem accounting for slight differences between energy densities in the hadronic 
(for baryons) and CS (for nuggets) phases and the presence of the propagating 
dark matter axions contributing to $\rho_{DM}$ see \cite{Ge:2017idw}. 

Unlike conventional dark matter candidates, such as Weakly Interacting Massive Particles 
(WIMPs) the AQN are strongly interacting and macroscopically 
large objects. However, they do not contradict the many known observational
constraints on dark matter or antimatter in the Universe for the following  
main reasons~\cite{Zhitnitsky:2006vt}: They carry a very large baryon charge 
$\la B \ra  \gtrsim 10^{25}$, and so their number density is very small $\sim B^{-1}$.  
As a result of interactions scaling with area, 
while number density is inversely proportional to the volume the small cross-section to 
mass ratio replaces the conventional WIMP requirement of weak interaction.
In other words, their interaction with visible matter is highly inefficient, and 
therefore, the nuggets are perfectly qualified  as  dark matter candidates.  
They do not contradict conventional  CMB results due to the same unique feature. 
Furthermore, 
the quark nuggets have a very large binding energy due to the large  gap 
$\Delta \sim 100$ MeV in the CS phases.  
Therefore, the baryon charge is so strongly bound in the core of the nugget that  
it is not available to participate in big bang nucleosynthesis
at $T \sim 100$~keV, long after the nuggets have been formed\footnote{\label{BBN}While the 
AQN model  basically does not modify the production of the dominant elements such as  hydrogen, 
helium and deuterium with $Z\leq 2$   as mentioned above, it may suppress the production of 
subdominant heavy nuclei with high $Z\geq 3$ due to the strong Boltzmann factor $\exp (Z)$ as argued 
in \cite{Flambaum:2018ohm}. Within AQN framework this represents a possible resolution of the 
primordial $^7Li$ puzzle in which conventional computations predict a higher value for the 
$^7Li$ abundance in comparison with observations.}. This large gap also 
prevents non-relativistic protons from easily penetrating into the nugget slowing the 
rate at which visible matter will annihilate with the antiquark nuggets at late cosmological 
times. Annihilation is a complicated process accompanied by a strong 
suppression factor which itself varies depending on the temperature, density 
and ionization of the surrounding environment in a non-trivial way. 

It should be noted that the galactic spectrum 
contains several excesses of diffuse emission the origin of which is unknown, the best 
known example being the strong galactic 511~keV line. The rare annihilation    
of the antimatter contained in an AQN with galactic visible matter could offer a 
potential source for several of these diffuse components
(including the 511 keV line \cite{Oaknin:2004mn, Zhitnitsky:2006tu} and accompanning  
continuum of $\gamma$ rays in the 100 keV to MeV range 
{\cite{Lawson:2007kp,Forbes:2009wg}, 
as well as x-rays \cite{Forbes:2006ba},  and radio frequency bands \cite{Forbes:2008uf}). 
It is important to emphasize that  all these emissions at drastically different frequencies 
scale with a single fundamental parameter of the model, the average 
baryon charge of the nugget $\la B\ra$, or equivalently, the axion mass $m_a$. 
This is because the rate of annihilation events  is proportional to 
one and the same product of the local visible and DM distributions at the annihilation site
defined as   
\be
\label{flux1}
\Phi \sim 4\pi R^2\int d\Omega dl [n_{\rm visible}(l)\cdot n_{DM}(l)],
\ee
where $R^2\sim B^{2/3}$ is the effective cross section $\sigma$  of interaction 
between the AQN and visible matter. As $n_{DM}\sim {\cal M}^{-1}\sim B^{-1}$ the effective 
interaction (\ref{flux1})  is strongly suppressed $\sim B^{-1/3}\ll 1$.   Precisely this 
small geometrical   factor replaces the conventional WIMPs requirement for   
weakness of the dark-visible coupling. The AQNs achieved this weakness 
not as a result of a new fundamental  weak coupling constant but rather through  
suppression as a result of large baryon number $B$ constrained by eq. (\ref{B-range}). 
   
To conclude our brief review of the AQN model we would like to mention the 
recent claim \cite{Zhitnitsky:2017rop,Raza:2018gpb,Zhitnitsky:2018mav}
that the AQNs  might be responsible for the resolution of  the renowned  (since 1939) 
``solar corona heating mystery".  The puzzle  is  that the corona has a temperature  
$T\simeq 10^6$K which is 100 times hotter than the surface temperature of the Sun, and 
conventional astrophysical sources fail to explain the extreme UV (EUV) and soft x ray radiation 
from the corona 2000 km  above the photosphere.   

It turns out that if one estimates the extra energy being injected  when the 
antiquark nuggets annihilate with the solar material one obtains
a total extra energy $\sim 10^{27}{\rm erg}/{\rm  s}$  which 
automatically  reproduces the observed EUV and soft x-ray energetics  \cite{Zhitnitsky:2017rop}. 
This estimate is derived  exclusively in terms of known  dark matter density 
$\rho_{\rm DM} \sim 0.3~ {\rm GeV cm^{-3}}$ and dark matter  velocity 
$v_{\rm DM}\sim 10^{-3}c $ surrounding the Sun without adjusting any  parameters 
of the model. This estimate is strongly supported by Monte Carlo numerical computations  
\cite{Raza:2018gpb} which suggest that most annihilation events occur precisely at the 
so-called transition region at an altitude of $2000$ km, where it is known that drastic changes 
in temperature and density occur.  

A ``smoking gun" supporting this proposal on the nature of the EUV  
would be the observation of axions radiated from the corona when the nuggets disintegrate 
in the Sun. The corresponding computations have been carried out recently in  
\cite{Fischer:2018niu,Liang:2018ecs}.  Presently the CAST   (CERN Axion Search Telescope) Collaboration  
has taken a significant step to upgrade its instrument to make it sensitive to the spectral 
features of the axions produced due to  the  AQN annihilation events  on   the Sun. 

In the context of the this work it is important to emphasize 
that this picture when the AQNs are responsible for the resolution of the 
``solar corona heating mystery" is consistent with the old idea advocated  by  Parker   
\cite{Parker} suggesting  that there must be small, sub-resolution events, the so-called 
``nanoflares", which uniformly heat the corona.
In most studies the term ``nanoflare" describes a generic burst-like event for any impulsive 
energy release on a small scale, without specifying its cause, the hydrodynamic consequences 
of impulsive heating are assumed without discussing their nature, see the
recent review papers \cite{Klimchuk:2005nx,Klimchuk:2017}.
The AQN framework offers a specific realization of these nanoflares in form of the AQN 
annihilation events, see  \cite{Zhitnitsky:2017rop,Raza:2018gpb, Zhitnitsky:2018mav} with details and 
references on the original literature on observations supporting this identification. 
This identification implies that the baryon number distribution must be in the range
\be
\label{B-range-flares}
10^{23}\leq |B|\leq 10^{28}.
\ee
It is a highly nontrivial consistency check for the proposal 
\cite{Zhitnitsky:2017rop, Raza:2018gpb,Zhitnitsky:2018mav} that the required window   
(\ref{B-range-flares}) is consistent with the range of mean baryon number allowed 
by the  axion and dark matter search constraints (\ref{B-range}) as these 
come from a number of different and independent  constraints extracted from astrophysical, 
cosmological, satellite and ground based observations. It is this range of AQN masses 
which will  be used in our estimates of their contribution to the background sky temperature 
in Section \ref{sky-temperature}. 

\section{\label{thermal}Thermal Radiation from Nuggets} 
The AQNs begin as cold objects as the initial heat of formation will be quickly released 
as radiation from the electrosphere as will be discussed below. The quark nuggets may be 
cold relative to the photon background because the electrosphere is characterized by 
a plasma frequency $(\omega_p)$ which prevents the photons from outside penetrating 
into the nuggets' interior and efficiently depositing heat \cite{Forbes:2008uf}.

While photon heating is very inefficient   
the antiquark nuggets are heated through interactions with the surrounding 
visible matter as a result of annihilation events in their interior.
This heating results in radiation which depends on environment: a denser environment
leads to more annihilations, a higher temperature and stronger radiation from the AQNs. Of 
particular importance is the fact that the nuggets emit a non-black body spectrum which 
will distort the low energy CMB spectrum\footnote{The spectrum of the nuggets is determined
by emission of photons in the electrosphere with temperature $T_N$ and deviates from black body 
radiation as discussed in \cite{Forbes:2008uf}. In particular, at at low energies the spectrum is 
logarithmic $dE/d\nu\sim \ln \nu$ in contrast with black body radiation  $dE/d\nu\sim  \nu^2$. 
Some additional constraints on the low energy tail emerge as a result of finite size effects  
of the electrosphere relative to the wavelength of the emitted photons.}. 
The spectrum of nuggets at low temperatures was analyzed in \cite{Forbes:2008uf} and 
was found to be,
\be
\label{eq:Nug_spec}
\frac{dE}{dt~d\nu ~dA} = \frac{32\pi^2}{45} \frac{\alpha^{5/2} (k_B T_N)^3}{(hc)^2} 
\sqrt[4]{\frac{k_B T_N}{m_e c^2}} \nonumber\\
\times \left(1+\frac{h\nu}{k_B T_N}\right) e^{-h\nu/k_b T_N} 
F\left(\frac{h\nu}{k_B T_N} \right) 
\ee
where $T_N$ is the nuggets' effective radiating temperature, $m_e$ is the electron mass 
and we have defined the function, 
\begin{eqnarray}
F(x) &=& 17 - 12 {\rm{ln}}\left(\frac{x}{2}\right) ~~~ x<1, \\
&=& 17 + 12 {\rm{ln}}\left( 2\right) ~~~ x>1 \nonumber 
\end{eqnarray}
to simplify notation. 
The spectrum (\ref{eq:Nug_spec}) is very hard at low frequencies $\nu\ll T_N$ relative 
to black body radiation which will play an important role in our analysis.  While it is not 
relevant at the energies considered here the spectrum in equation (\ref{eq:Nug_spec}) has 
a low energy cutoff due to finite size effects within the electrosphere. As discussed in 
\cite{Lawson:2012zu} this effect strongly suppresses the present day radio 
background below $\sim 10$MHz.
\exclude{\footnote{As the resulting spectrum is strongly peaked at 
low energies this cutoff is important in setting the total energy required to heat the low energy 
photons above the CMB temperature. However, the exact details of the cutoff are not relevant 
to the present studies as the required number of baryon annihilations remains small for any 
reasonable low energy cutoff.}.}

The total energy emitted by the nugget per unit time per unit area $dA$ is given by 
integrating expression (\ref{eq:Nug_spec}) over all emission frequencies to obtain, 
\begin{equation}
\label{eq:total_emiss}
\frac{dE}{dt~dA} = \frac{128 \pi^2}{3} \frac{\alpha^{5/2} (k_B T_N)^4}{h^3c^2} 
\sqrt[4]{\frac{k_B T_N}{m_e c^2}}.
\end{equation}
The corresponding intensity is almost 6 orders of magnitude smaller than 
black body radiation for temperatures at the eV scale.
The nuggets are therefore very inefficient emitters at the time of CMB formation and 
do not distort the conventional CMB results\footnote{Nevertheless, this rate of emission is 
fast compared with cooling rate of CMB radiation due to the Universe's expansion and  
the nuggets' temperature would drop very fast, in a matter of seconds,  if annihilation events 
heating the nuggets were not happening.}. However, the low energy tail (which scales as 
$\sim \ln \nu$) of the emission from the nuggets is drastically different from black body radiation 
$\nu^2$. This observation will play a key role in our arguments which follow.  

As argued above the AQN are not effectively heated by the relatively low energy photons of 
the radiation background. As such the nuggets composed of quarks will have a temperature 
well below the radiation. This is not however the case for nuggets composed of antiquarks which may 
be heated by the capture and annihilation of of surrounding matter. If the antiquark nuggets interact 
sufficiently to be in radiative equilibrium the energy output of expression (\ref{eq:total_emiss}) 
must be balanced by the rate at which annihilations deposit energy in the AQN\footnote{While 
the majority of energy deposited in the nugget fully thermalizes some small amount will be emitted 
at higher energies. The resulting cosmic background also produces a constraint on the AQN model 
but it is considerably weaker than the effects discussed here. See appendix \ref{high-energy-radiation}
for further details.}. In particular the energy flux onto an AQN should satisfy, 
\begin{equation} 
\label{eq:matter_flux}
\frac{dE}{dt~dA} \propto \rho_{\rm vis} v
\end{equation} 
such that the nuggets will be warmer in environments with larger visible matter densities 
and temperatures. If the surrounding environment is sufficiently ionized the energy input may 
differ from the basic expression (\ref{eq:matter_flux}) due to the influence of long range 
electromagnetic interactions. 

The total amount of energy injected into the plasma is negligibly small and does not modify 
the conventional cosmological analysis, as argued in the simple estimates in  \ref{pre-BBN}.  
The only comments we will make here is that the observational  consequences advocated in  
this work emerge due to a very specific feature of the low energy  spectrum: it 
scales as $\ln \nu$  in eq. (\ref{eq:Nug_spec}) instead of the conventional $\nu^3$ for  
radiation of  a plasma at temperature $T$. The corresponding portion of the spectrum generates  a 
negligible amount  of energy at the moment of emission as stated above.   However, this excess 
becomes important at later times as it is redshifted across the 21 cm transition as 
a result of expansion as will be discussed in section \ref{sky-temperature}. 

The radiating temperature of the nuggets will evolve with redshift following a power law 
of the form  
\begin{equation}
\label{eq:temp_evo}
T(z) = T_{\rm rec} \left( \frac{1+z}{1+z_{\rm rec}} \right)^\beta 
\end{equation}
where $z_{\rm rec} \approx 1100$ is the redshift at recombination and $T_{\rm rec}$ 
is the nuggets'  average radiative temperature at that time. 
We may estimate the power law exponent $\beta$ 
by noting that the total thermal emission in expression (\ref{eq:total_emiss}) scales as 
$T_N^{17/4}$. If the heating is annihilation dominated then the energy 
input follows equation (\ref{eq:matter_flux}) which scales as $(1+z)^{3.5}$ if the matter 
is thermally coupled to the CMB or as $(1+z)^{4}$ if it is cooling adiabatically. These two 
possible scaling laws would imply either $\beta \approx \frac{14}{17}$ or 
$\beta \approx \frac{16}{17}$. It should be noted that emission is strongly dominated by 
the contribution from large $z$ so that the observable spectrum is largely insensitive to the 
precise value of $\beta$. More important is the nuggets' effective radiating temperature 
at the time the universe first becomes transparent to low energy radiation.

If the nuggets are in radiative equilibrium with the surrounding environment then we expect 
the radiative output of equation (\ref{eq:total_emiss}) must balanced the rate at which annihilations 
deposit energy in the nugget according to expression (\ref{eq:matter_flux}). 
In this case we may write, 
\begin{equation}
\label{eq:rad_balance}
\frac{128 \pi^2}{3} \frac{\alpha^{5/2} (k_B T_N)^4}{h^3c^2} \sqrt[4]{\frac{k_B T_N}{m_e c^2}}
= \kappa \rho_{\rm vis} v
\end{equation} 
The factor $\kappa$ is introduced to account for the fact that collision rate need not strictly match 
the annihilation rate as not all matter striking the nugget will 
annihilate and not all of the energy released by an annihilation will be thermalized in the nuggets.
As such $\kappa$ encodes a large number of complex processes including the probability that 
low momentum protons are unlikely to overcome the larger gap and penetrate into the 
colour superconducting phase of the nugget 
as well as charge exchange effects between the hydrogen atom and the nugget. These may involve 
either the annihilation of the electron or charge exchange between the proton and the central quark 
matter or interactions with a hole state in the electrosphere. In a neutral environment when no 
long range interactions exist the value of $\kappa$ cannot exceed $\kappa \sim 1$ which would 
correspond to the instantaneous and total annihilation of all impacting matter into to thermal 
photons. The high probability of reflection at the sharp quark matter surface lowers the value of 
$\kappa$ while ionization due to electron annihilation increases the value of  $\kappa$. 
 
Annihilation events are suppressed by a number of complicated many-body effects. In  
particular, an incoming proton for a successful annihilation must overcome a sharp 
boundary between the CS and hadronic phases. Also,   
a successful annihilation is accomplished only if the quantum numbers of the three quarks of the 
proton exactly match  the quantum numbers of a drastically different di-quark condensate and a 
quark excitation in CS phase. In addition to these primary suppression mechanisms there are many other
many-body effects which must be taken into account for a proper estimation  of parameter $\kappa$. 
For that reasons a computations of the parameter $\kappa <1$ from  first principles 
in strongly coupled QCD is not even feasible at the moment. 
A determination of $\kappa$ also depends on the environment and includes 
accounting for the fraction of annihilation energy lost through 
non-thermal emission\footnote{Near surface annihilations may result in the emission of both x-rays 
\cite{Forbes:2008uf} and charged particles \cite{Lawson:2010uz}. However the branching fraction to 
these higher energy products is expected to be small and is constrained by both the cosmic background 
as discussed in appendix \ref{high-energy-radiation} and by present day 
observations\cite{Lawson:2015cla}.}. Given the impracticality of directly calculating the value of 
$\kappa$ we will adopt a phenomenological approach in which we determine the value of 
$\kappa$ at $z\simeq 17$ consistent with the EDGES observation and then comment on the 
consistency of this value with physical expectations and with other observational constraints on 
the model. 

We start our task by estimation of the average velocity of the hydrogen.   
At the time of recombination the photons have a temperature of $\approx 0.3~ {\rm eV}$ and the neutral 
hydrogen is expected to follow a Maxwell distribution at this temperature. Consequently the average 
velocity of the hydrogen is,
\begin{equation}
\label{v}
\la v \ra = \sqrt{\frac{8k_BT_{\gamma}}{\pi m_p}} \approx  9\times 10^5 {\rm{cm}}/{\rm{s}}.
\end{equation}
The hydrogen is also very uniformly distributed with a density,
\begin{equation}
\label{rho}
\rho_{\rm vis} = \Omega_B \rho_c (1+z_{\rm rec})^3 \approx 330~ \frac{m_pc^2}{{\rm{cm}}^3},
\end{equation}
where $\Omega_B$ is the baryon fraction, $\rho_c$ is the critical density and $m_p$ is the proton mass. 
Substituting (\ref{v}) and (\ref{rho}) to (\ref{eq:rad_balance}) we arrive to our estimate for the 
temperature in terms of unknown parameter $\kappa$, 
\begin{equation}
\label{eq:TR}
T_{\rm rec} = 0.7~{\rm{eV}}\cdot  \kappa^{4/17}. 
\end{equation} 
It is instructive to compare the obtained estimate (\ref{eq:TR})  with the analogous 
study of the galactic environment  \cite{Forbes:2006ba,Forbes:2008uf}
where the internal  temperature of a nugget was estimated on the level $T\simeq 1$ eV.
In both cases the numerical values for the densities  $\rho_{\rm vis}$ assume  the same order 
of magnitude, while the galactic velocity
$\sim 10^{-3}c$ is slightly higher than (\ref{v}). This leads to numerical similarity 
between $T_{\rm rec}$ given by (\ref{eq:TR}) and $T\simeq 1$ eV  obtained for the galactic 
environment \cite{Forbes:2006ba,Forbes:2008uf}. 

It should also be noted that while expression (\ref{eq:TR}) 
may allow for the AQN to have a temperature comparable 
to that of the photons they have little impact on the present day CMB as any effects are suppressed 
by both the baryon to photon ratio (expressed in terms of $\eta\sim 10^{-10}$) and the 
small cross section to mass ratio of the AQN 
(represented by parameter $\sigma/M \sim B^{-1/3}$), similar to our estimates of the AQN 
impact on pre-BBN effects carried out in  \ref{pre-BBN}. 

We emphasize that the key ingredient  generating  the observable effect from AQN   
is the drastic difference in the spectrum in equation (\ref{eq:Nug_spec}) with its $\ln \nu$ 
behaviour  in comparison with the conventional black body portion   $\nu^2$ of the CMB at low 
frequencies. These distinct spectral features will play the crucial role in our arguments in the next section. 

\section{\label{sky-temperature}Soft Photon Excess from AQN  Dark Matter}
As was claimed  in \cite{Feng:2018rje} a stronger than predicted radiation background 
present before $z\sim20$ may produce the strong 21 cm absorption feature 
observed by EDGES. This results from the larger misalignment between the radiation 
temperature and the spin temperature of neutral hydrogen 
with the strength of this feature scaling as,
\begin{equation}
\label{eq:absorb_scale}
\delta T \sim   x_{HI} \left( 1 - \frac{T_r}{T_s} \right) 
\end{equation}
where $x_{HI}$ is the fraction of neutral hydrogen and $T_r$ and $T_s$ are the 
radiation and gas spin temperatures respectively. 

In what follows we will  demonstrate that the radiation background  
produced by the nuggets mimics that required to produce the 21 cm absorption 
strength according to the computations of ref. \cite{Feng:2018rje}. To be more specific,  
we want to reproduce the power law spectrum observed today
\begin{equation}
\label{eq:ARCADEfit}
T(\nu) = T_{CMB} + \xi T_R \left( \frac{\nu}{\nu_0} \right)^{\beta}
\end{equation} 
with the parameters $T_{CMB} = 2.729$K, $\xi \approx 0.01, T_R = 1.19$K, 
$\nu_0 = 1$GHz, $\beta = -2.62$ which \cite{Feng:2018rje} claims will reproduce 
the EDGES result provided the radiation is generated before $z\sim 20$. 

In order to estimate the present day background contribution from the nuggets 
we need to integrate their contribution over all redshifts from recombination down 
to $z\approx 20$. While the nuggets will continue to produce radiation below $z=20$ 
it is only their large $z$ contribution which contributes to 21 cm absorption and 
should be compared to the spectrum in expression (\ref{eq:ARCADEfit}).  This 
contribution was previously estimated in \cite{Lawson:2012zu} and is given by the integral, 
\begin{equation}
\label{eq:z_int}
I(\nu) = \int_{20}^{1100} \frac{2 c dz}{3 H(z) (1+z)} 
\rho_{DM} \la \frac{\sigma}{\cal{M}} \ra \frac{dE}{d\nu~dt~dA}  
\end{equation}
where $\la{\frac{\sigma}{\cal{M}}}\ra\sim \la B\ra^{-1/3}$ is the average 
cross-section to mass ratio of the nuggets
and $\frac{dE}{d\nu~dt~dA}$ is the emission spectrum given in equation 
(\ref{eq:Nug_spec}) evaluated at the redshift corrected frequency $\nu(1+z)$ 
and at the nugget temperature determined at a given redshift by expression (\ref{eq:temp_evo}). 
The factor of $2/3$ accounts for the fact that only the antiquark nuggets from 
(\ref{ratio}) will be sufficiently heated to contribute to the spectrum at MHz frequencies. 
The intensity calculated in expression (\ref{eq:z_int}) is dependent on both the average 
baryon number $\la B\ra$ of the AQN population  and the nugget temperature.  However
the resulting expression scales with the nuggets' average cross-section to mass ratio and is 
thus suppressed at large baryon number as $B^{-1/3}$ so that the value of $T_{\rm rec}$ 
required to match the spectrum (\ref{eq:ARCADEfit}) remains at the fraction of an eV 
scale within the allowed window  for the baryon charge (\ref{B-range}) for the nuggets.

For a given average baryon number we can solve for the $T_{\rm rec}$ value that reproduces 
the strength of the model (\ref{eq:ARCADEfit}) at the centre of the 21cm feature at 
$\nu \approx 75 MHz$. The required $T_{\rm rec}$ as a function of baryon number is shown 
in Figure \ref{fig:T_LS_B} and falls in the sub eV range across a wide range of nugget 
sizes. As a larger baryon number results in a smaller cross-section to mass ratio 
heavier nugget populations require a higher initial temperature 
to produce the radiation field suggested by the EDGES effect. As the maximum predicted 
values of $T_{\rm rec}$ is on the order of 0.7eV according to (\ref{eq:TR}) 
the EDGES effect can only be fully saturated by nuggets 
with an average baryon number below $\la B\ra \lesssim10^{29}$. Larger $\la B\ra $ 
values are not excluded by 
the EDGES result, but are insufficient to fully explain the observed feature. One should emphasize that
this result is consistent with the window (\ref{B-range-flares}) derived from  
completely independent observations.

As may be seen in Fig.
\ref{fig:T_LS_B} matching the required sky temperature demands that the nuggets have a 
radiating temperature slightly below the eV scale when the universe first becomes 
transparent to low energy photons. As argued at the end of section \ref{thermal} a precise 
estimate of $T_{\rm rec}$ is a complicated problem of the non-perturbative QCD,
and highly sensitive to the microscopic details. 
However, we note that the lower bound on average AQN size (which is around $\la B\ra \sim 10^{25}$ 
as reviewed in Sect.\ref{sec:AQN}) corresponds to $T_{\rm rec} \sim 0.3$ eV. According to 
expression (\ref{eq:TR}) this corresponds to a 
lower bound on the annihilation efficiency of $\kappa \gtrsim 0.03$.  

\begin{figure}
\includegraphics[width=\linewidth]{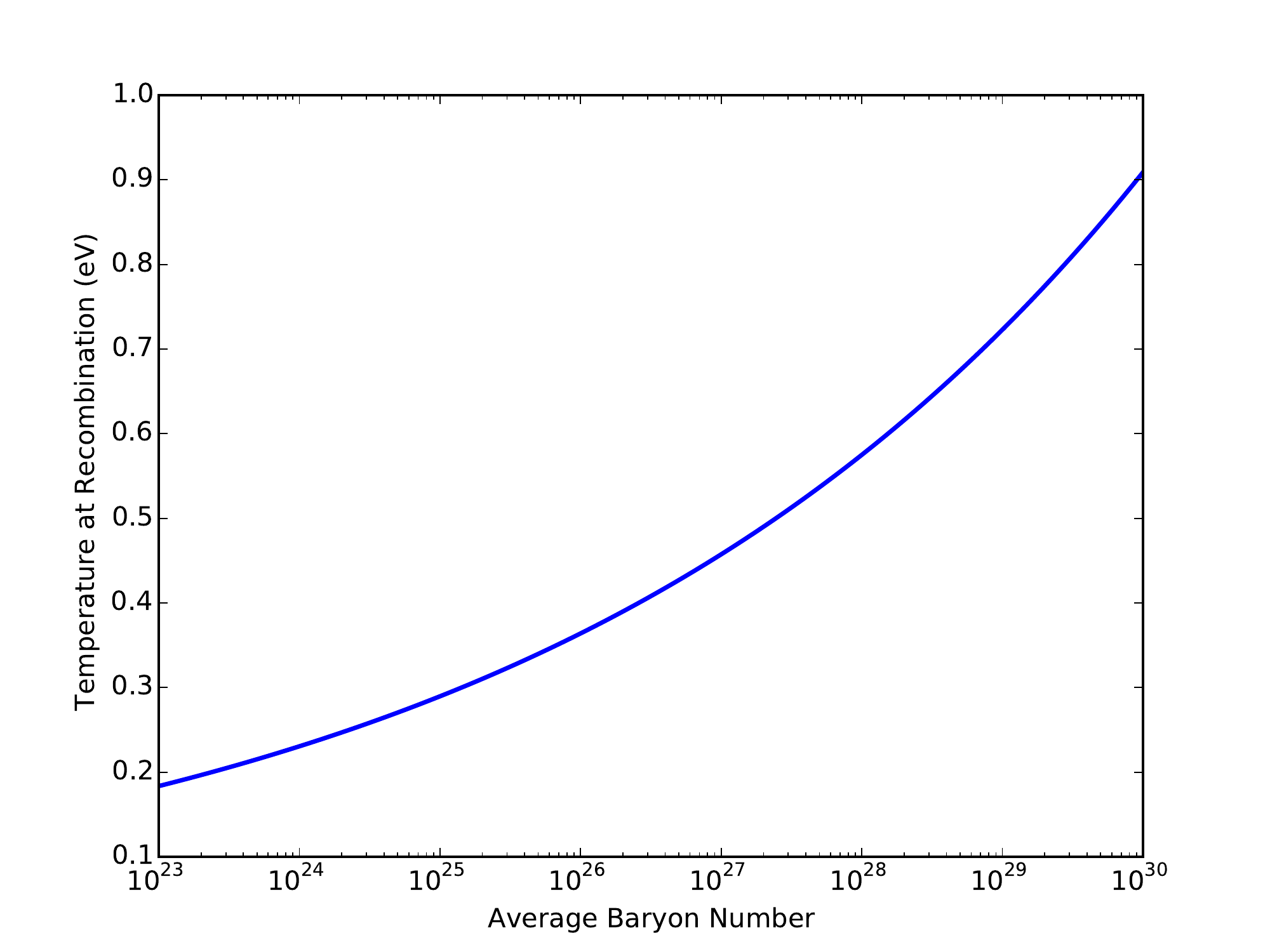}
\caption{Nugget temperature required to produce $1\%$ of the ARCADE2 excess 
as measured at $\nu = 75$MHz as a function of the nuggets mean baryon number.}
\label{fig:T_LS_B}
\end{figure}

A few comments on the allowed range $1>\kappa \gtrsim 0.03$ are in order. As argued above
the value of $\kappa$ is determined by a balance between the large probability of reflection at the 
quark nugget surface and the tendency of neutral hydrogen to loose its electron and become bound 
to the nugget. The case $\kappa=1$ corresponds to a situation in which capture is completely 
efficient and all protons hitting the nugget annihilate. As explained above the probability 
for successful annihilation is expected to be strongly 
suppressed as it requires a perfect overlap of the wave-functions between  the 
non-hadronic states of the colour superconductor and  quarks from the proton. 
Within this context a suppression factor at the $\kappa \sim 0.1$ scale seems well justified, and 
in fact assumes a similar numerical value $\kappa_{\rm galactic}\sim 0.1$ extracted from the 
analysis of the galactic radiation excess in \cite{Forbes:2008uf, Forbes:2006ba}, and we want to 
elaborate on the nature of this similarity now. The  visible matter flux onto the nuggets within 
the galactic centre environment is similar to that discussed above at the time of recombination. 
Following \cite{Forbes:2006ba,Forbes:2008uf,Lawson:2012zu} 
we may estimate the matter flux in the galactic centre as,
\begin{equation}
\label{eq:gcflux}
\rho_{\rm gc} \la v_{\rm g}\ra \approx  2\times 10^9 \frac{\rm GeV}{\rm{cm^2 s}} 
\approx 7 \rho_{\rm vis} \la v \ra  |_{z=1100},
\end{equation}
where the estimate on the right hand side is based on equations (\ref{v}) and (\ref{rho}). 
This similarity in fluxes (\ref{eq:gcflux}) for two different environments can be directly translated into 
a similarity in the AQN temperatures for these two cases.  An increase of the flux by factor $7$ or so
translates to a modest increase of the temperature (which scales as $T^{17/4}$ according to 
(\ref{eq:rad_balance}))  by  factor of $\sim 1.6$ in comparison with our estimate (\ref{eq:TR}). 

These estimates using simple rescaling arguments should not be considered as solid predictions as 
there are a number of complications in the galactic environment. In particular, the galactic centre 
population may experience a further boost in heating due to the increased scattering cross section  
due to a substantial ionization fraction.  The analysis of \cite{Forbes:2008uf} found that unaccounted for 
diffuse radio emission from the galactic centre was saturated by the AQN contribution when $T\sim$eV 
and that the global fit to galactic emissions could be improved by adding an AQN source near this scale. 
While subject to large uncertainties in both the dark and visible matter distribution in the galactic centre 
as well as the contribution of other diffuse radio sources this result is consistent with our estimate 
of the value of $T_{\rm rec}$ required to produce the 21cm feature observed by EDGES arising  
at a radically different epoch. 

As mentioned previously, our goal is not a computation of $T_{\rm rec}$, which would be an 
extremely difficult task involving many unknowns in strongly coupled QCD. Instead, our goal 
is to argue that the required value of $T_{\rm rec}$ which may explain the EDGES result is perfectly 
consistent with our previous galactic computations and with our broader 
understanding of how the AQN will behave in different environments. As a further consistency 
check we note that the estimated $T_{\rm rec} \lesssim 0.7$eV range (as shown in figure \ref{fig:T_LS_B}) 
corresponds to an average baryon number which is perfectly consistent with 
the window (\ref{B-range-flares}) for $\la B\ra $ derived from completely independent observations. 

The spectrum which results from performing the integration of expression (\ref{eq:z_int})
is shown in Fig. \ref{fig:SkyTempB26} with the temperature $T_{\rm rec} \approx 0.3$ eV
chosen to match the required sky temperature at $\nu = 75$MHz. The key observation 
here is that  the spectrum  (\ref{eq:Nug_spec}) is almost constant for small frequencies 
at $\nu\ll T_{\rm rec} $, while the CMB has conventional black-body suppression at small $\nu$. 
This is the technical explanation why the spectrum  (\ref{eq:Nug_spec}) scales 
similarly to (\ref{eq:ARCADEfit}) when expressed in terms of the sky temperature. 

\begin{figure}
\includegraphics[width=\linewidth]{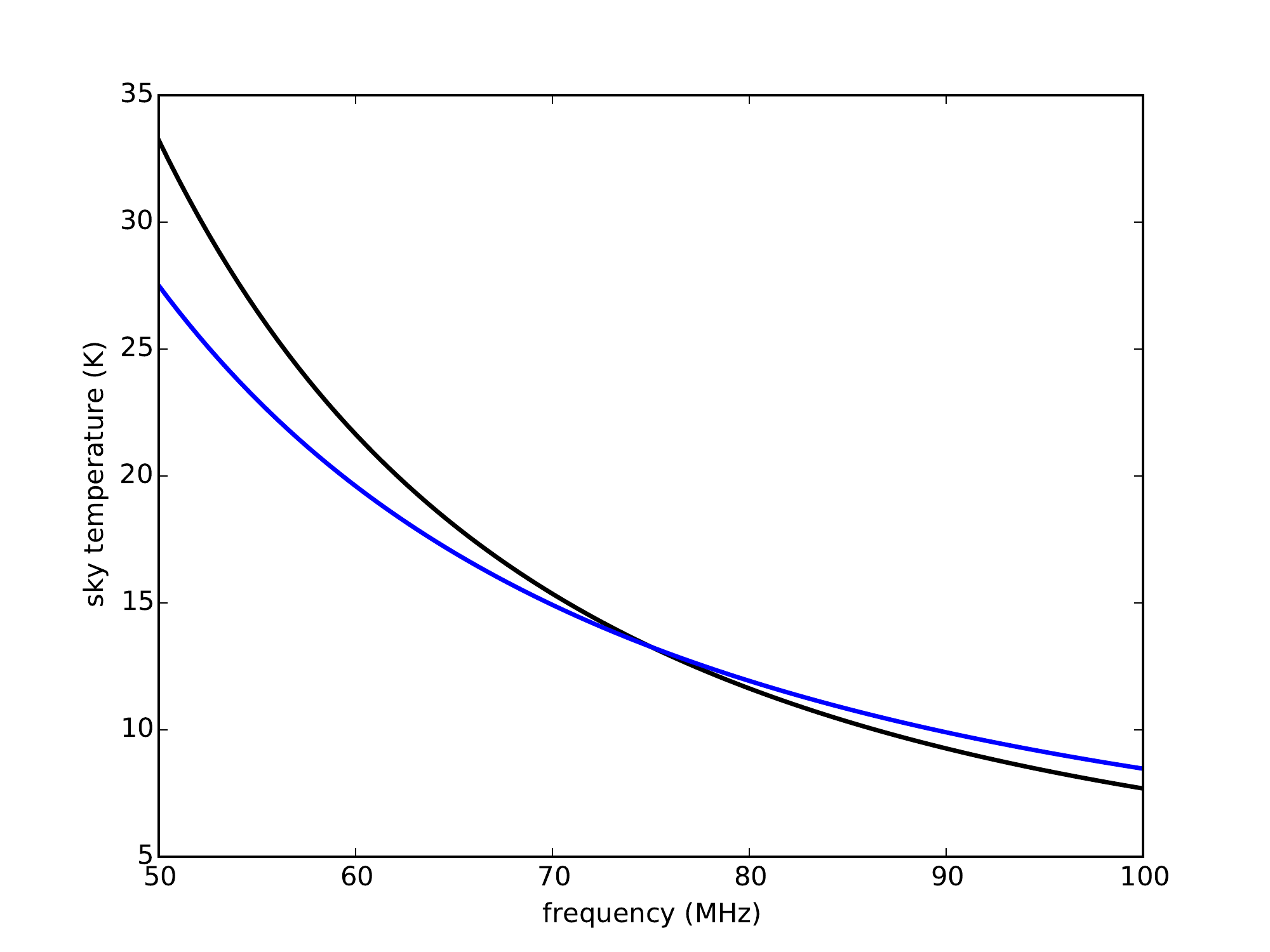}
\caption{Sky Temperature as a function of frequency across the EDGES low-band 
frequency range (50-100MHz). The approximate  temperature profile required to reproduce 
the observed 21 cm absorption feature is shown in  black as is the contribution 
from nuggets of size $B=10^{26}$ with the initial temperature $T_{rec} \approx 0.3$eV 
chosen to match the the required temperature excess 75MHz. 
The analysis of ref. \cite{Feng:2018rje} was based on formula (\ref{eq:ARCADEfit})
which may be reproduced by the AQN model with accuracy better than $15\%$  
across the entire relevant region.}
\label{fig:SkyTempB26}
\end{figure} 

While an additional large redshift source of of radio band emission may ease the 
tension between present cosmological models and the EDGES result the same 
result limits the ionizing radiation contribution of any such source. As may be seen
from the scaling in expression (\ref{eq:absorb_scale}) the absorption feature will 
be strengthened only if the increase in radiation temperature is not accompanied by 
a similar increase in the ionization fraction or gas temperature. In 
\ref{high-energy-radiation} we present simple  arguments suggesting that 
composite dark matter models (such as the AQN model advocated in the present work) 
are uniquely able to provide a radiation background 
of this type due to the necessary presence of dark matter at early times 
and the ability of composite objects to thermalize 
any energy produced. In conventional dark matter models (formulated in terms of 
local weakly interacting fields, such as WIMP based models) the dark matter particles  
typically produce high energy photons or other ionizing particles which can generate 
a radio band signal only through subsequent interactions with the surrounding 
visible matter and which may significantly alter the ionization fraction.  
 
As may be seen in Fig. \ref{fig:SkyTempB26} the spectrum of the nuggets can 
closely match the required radio excess with a suitable choice of 
initial temperature for any given average  baryon number within  the window (\ref{B-range-flares}). 
This is the main result of the present work. Essentially, by explicit computations within 
the AQN framework,  we support the claim made in ref.\cite{Feng:2018rje} that a $\sim 1\%$ 
contribution to ARCADE2 excess from radiation at earlier times may explain the EDGES result.
Our contribution into the field is modest: we demonstrated  that the AQN model can naturally provide 
the required excess without conflicting with numerous strong  constraints which other DM proposals 
face as overviewed in the Introduction.

\section{Conclusion}
We have argued that AQN dark matter will produce an excess of soft 
photons at early times. If these photons sufficiently increase the radiation temperature 
at long wavelengths they may produce a larger than anticipated difference between the 
radiation temperature and the spin temperature of the neutral hydrogen gas. This 
in turn, as argued in  \cite{Feng:2018rje}, will account for the stronger than 
anticipated 21 cm absorption feature centred at $z\approx 17$ observed by EDGES. 
Our computations in the AQN model should be considered as 
an explicit realization of the idea formulated in ref.\cite{Feng:2018rje} that a 
$\sim 1\%$ contribution to the ARCADE2 excess from radiation present at $z\sim 17$ 
may explain the EDGES result.

Our claim is that the AQN dark matter model is well suited to 
providing such a soft photon excess as the nuggets have a large number of low energy 
excitation levels (associated with positrons far from the nugget's surface) which will 
dominate their emission spectrum. In this sense the EDGES result, if confirmed, would 
seem to favour composite dark matter models in which any energy release may be 
thermalized down to the lowest available emission modes. The same composite structure 
of the nuggets prevents equilibration of the CMB photons with the internal temperature 
of the nuggets. 

The technical reason for this effect  to emerge is related to a very specific and unique features 
of the low energy  spectrum: it scales as $\ln \nu$  in eq. (\ref{eq:Nug_spec}) instead of  $\nu^3$ 
for conventional radiation from a plasma at temperature $T$.
This portion of the spectrum contributes very little to the total photon energy density. But this low 
energy portion of the spectrum could be important as a result of the Universe expansion at later times, 
as argued in present work. 

We reiterate that this AQN model was invented as the natural explanation of the observed 
ratio (\ref{eq:dark-vis}), suggesting that the dark matter $ \Omega_{\rm dark}$ and the visible 
matter $\Omega_{\rm visible}$ formed at the same cosmic epoch. Some of the history 
of this model is traced in Section \ref{sec:AQN} and the references therein. This should be
contrasted with a large number of recently introduced models specifically designed to explain 
the EDGES result. 

\section*{Acknowledgements} 
 This research was supported in part by the Natural Sciences and Engineering 
Research Council of Canada.

\appendix
\section{\label{high-energy-radiation}Higher Energy Radiation}
In this Appendix we argue that the high energy radiation which necessarily accompanies   
the radio band emission is negligible in the framework of the AQN dark matter model. 

In particular the major portion of the energy resulting from the annihilation of a nucleus 
in the bulk of an AQN will be thermalized while roughly half of the near surface electron 
annihilations will result in the emission of a $\gamma$-ray with energy of order $m_e$. 
We may thus estimate the relative strength of thermal emission to ionizing high energy 
radiation as $m_p/m_e \approx 2000$.  

An order of magnitude estimate of the HI ionization fraction resulting due to 
$\gamma$-ray emission from the nuggets demonstrates their minimal impact 
on gas dynamics. If we assume that every collision of an atom with an antiquark 
nugget results in the emission of a $\gamma$ ray of average energy $E\sim m_e c^2$ 
then the maximum rate at which a nugget can ionize the surrounding gas is, 
\begin{equation}
\frac{dn_{ion}}{dt} \approx \frac{m_e c^2}{I_{HI}} \frac{\rho_{B}}{m_p} v_B \sigma_N
\end{equation}
where $I_{HI} = 13.6$eV is the ionization energy of hydrogen. Note that, as discussed 
in the case of nuclear annihilations, it is likely that only a fraction of all collisions will 
result in an electron annihilation. This estimate is therefore an upper bound on ionizing 
radiation assuming all incident atoms result in a high energy photon being emitted. 
Scaling this expression by the antiquark nugget to hydrogen ratio,
\begin{equation}
\frac{n_N}{n_{HI}} = \frac{ \Omega_{DM} m_p}{ \Omega_{B} M_N}
\end{equation}
gives the rate of change of the ionization fraction,
\begin{equation}
\frac{dx_{HI}}{dt} \approx \frac{ \Omega_{DM}}{ \Omega_{B} M_N}
\frac{m_e c^2}{I_{HI}} \rho_{B} v_B \sigma_N.
\end{equation}
Finally noting that $t \approx H^{-1}$ allows us to integrate this expression over redshift. 
\begin{eqnarray}
&&\Delta x_{HI} = \int dz \frac{dt}{dz} \frac{dx_{HI}}{dt} \nonumber \\
&\approx& \frac{\rho_c \Omega_{DM}}{H_0\sqrt{\Omega_M}} \frac{m_e c^2}{I_{HI}} 
v_0 \frac{\sigma_N}{M_N} (1+z_{LS})^2
\end{eqnarray}
where $v_0 \sim 150$m/s is the hydrogen atom velocity scale at 
$T_{CBM}$ and we have used the fact that emission is strongly peaked at early 
times when the universe is matter dominated. 
Finally, we may formulate this in terms of the average baryon number of the nuggets,
\begin{equation}
\Delta x_{HI} \approx 3\times 10^{-10} \left( \frac{10^{26}}{B}\right)^{1/3}.
\end{equation}
Which is a negligibly small change in the ionization fraction. 

Emission from the nuggets at energies $E \sim m_e c^2$ produced at $z=1100$ will 
be redshifted down to the keV scale and are thus sensitive to the limits from 
the {\textit{Chandra}} x-ray background observations \cite{Cappelluti:2017miu}. 
However the majority of these photons are absorbed and ionize the neutral hydrogen 
as discussed above and the remaining x-rays  are well below the observed backgrounds.

\section{pre-BBN cosmology: AQN annihilation events  and energy injection}\label{pre-BBN} 
In this Appendix we provide a few estimates related to energy injection during the pre-BBN cosmology
when $T> 1$ MeV and the density of electrons and positrons is very high with $n_e \sim T^3$.
Naively one might think that the large density of the plasma (and, therefore, frequent  
interactions with the AQNs) could drastically modify the standard pre-BBN cosmology. 
Nevertheless, as we argue below, the main conclusion is that the corresponding rate of energy 
injection is negligible and does not modify the standard pre-BBN cosmology.

The basic reason for this conclusion is that the number density of the nuggets 
$n_{\rm AQN} \sim B^{-1}$ is very tiny relative to the photon density due to the large baryon charge 
of the nuggets $B\sim 10^{25}$ and the small baryon to photon ratio $\eta\sim 10^{-10}$. 
The small energy injection rate, to be estimated below, is a direct consequence of this suppression factor. 

We start by estimating the number of annihilation events per unit time for a single nugget. At $T>1$MeV
annihilations are dominated by the thermally abundant electrons and positrons which may annihilate 
in the electrosphere bound to the AQNs. These collisions occur at a rate,
\be
\label{1}
\frac{dN}{dt}\sim 4\pi R^2 n_{\rm e} c
\ee
where number density of electrons, positrons  and photons  is estimated as 
$n_{\rm e}\sim n_{\rm e^+}\sim n_{\gamma}\sim T^3$. 
The energy injection per unit time for a single nugget can be estimated from (\ref{1}) by 
multiplying a typical  energy being produced as a result of annihilation:
\be
\label{2}
T\frac{dN}{dt}\sim 4\pi R^2 T n_{\rm e} c.
\ee
The energy injection per unit volume per unit time can be estimated as
\be
\label{3}
\frac{dE}{dVdt}\sim 4\pi R^2 T n_{\rm AQN} n_{\rm e}c, 
\ee
where the AQN number density $n_{\rm AQN} $ is estimated as follows 
\be
\label{4} 
n_{\rm AQN}\sim \frac{n_B}{\la B\ra} \sim \eta \frac{n_{\gamma}}{\la B\ra} \sim \frac{T^3 \eta}{\la B\ra}\sim 10^{-35}T^3 ,
\ee
where $ \eta\sim 10^{-10}$ is the baryon to photon ratio. We want to estimate the total amount of 
energy injected into the system per unit volume during a mean free time $\tau\sim (\alpha^{2}T)^{-1}$ 
which is defined as a typical time between collisions. The timescale $\tau$ is representative of the 
time needed for the system to adjust to the injected energy and return to equilibrium. We want to 
compare this extra injected energy (due to the AQNs) with the typical energy density in the system, 
i.e.we consider the dimensionless ratio
\be
\label{5}
\frac{1}{(T n_{\rm e})}\frac{dE}{dV}\sim    \frac{\left(R^2 T^2\right)\eta}{\alpha^{2} \la B\ra} 
\sim 10^{-19}\left(\frac{T}{1 \rm MeV}\right)^2.
\ee
It is clear that the small amount of energy injected into the system is quickly equilibrated within 
the system so that the standard pre-BBN cosmology remains intact.  In other words, the conventional 
equation of state, and conventional evolution of the system are unaffected  by presence of AQNs. 

The annihilation processes of the AQNs with protons produce even smaller effect due to the 
additional suppression $n_B/n_e\sim \eta$ as the baryon  density is drastically smaller than the 
electron density at $T> 1 $ MeV.   The larger amount of energy released per collision 
$\sim m_p/T$ cannot overcome  an additional suppression due to the very small factor $\sim \eta$. 
This argument, in fact, holds for much lower temperatures $T\approx 20 $ keV when the number densities  of protons and electrons
become approximately the same order of magnitude \cite{Flambaum:2018ohm}.

The simple estimate presented above shows  that the interaction of the AQNs with surrounding 
material is strongly suppressed due to the $B^{-1}$ factor appearing in the interaction rate. 
The observational consequences advocated in the main body of this work emerge exclusively 
due to very specific and unique  features of the low energy spectrum which scales as $\ln \nu$ 
instead of conventional $\nu^2$ seen in the low energy tail of a blackbody spectrum.
This low energy portion of the spectrum contains too little energy to effect the physics of the 
early universe. However, emission produced after recombination will contribute to the CMB producing 
a warmer than blackbody CMB spectrum at low frequencies.

\end{document}